\documentclass[aps,10pt,floatfix,twocolumn,showpacs]{revtex4-1} 
\usepackage{color,soul}
\usepackage{graphicx}
\usepackage{siunitx}
\usepackage[breaklinks=true,colorlinks=true,linkcolor=blue,urlcolor=blue,citecolor=blue]{hyperref}

\usepackage{dblfloatfix}
\usepackage{amsmath, amsthm, amssymb}
\usepackage{amsthm}
\usepackage{multirow}
\usepackage[normalem]{ulem}
\usepackage{soul}
\usepackage{makecell}
\usepackage[table]{xcolor}
\definecolor{Gray}{gray}{0.85}
\definecolor{LightCyan}{rgb}{0.88, 1, 1}
\definecolor{Apricot}{rgb}{0.98, 0.81, 0.69}
\usepackage{threeparttable}
\newcommand{\be}{\begin{equation}}
\newcommand{\ee}{\end{equation}}
\newcommand{\bea}{\begin{eqnarray}}
\newcommand{\eea}{\end{eqnarray}}

\usepackage[rightcaption,]{sidecap}
\sidecaptionvpos{figure}{c}
\usepackage[utf8]{inputenc}
\usepackage{amsmath}
\usepackage{amssymb}
\usepackage{graphicx}
\usepackage{hyperref}
\usepackage[export]{adjustbox}
\usepackage[update]{epstopdf}
\usepackage{multirow}
\usepackage{placeins}

\begin{document}
\title{Accelerated Collapse Kinetics of Charged Polymers in Good Solvent: Role of Counterion Condensation}
\author{Susmita Ghosh}
\email{sghosh@thphy.uni-duesseldorf.de}
\affiliation{Institut für Theoretische Physik II - Soft Matter, Heinrich-Heine-Universität Düsseldorf}

\author{Satyavani Vemparala}
\email{vani@imsc.res.in}
\affiliation{The Institute of Mathematical Sciences, C.I.T. Campus,
Taramani, Chennai 600113, India}
\affiliation{Homi Bhabha National Institute, Training School Complex, Anushakti Nagar, Mumbai, 400094, India}

\date{\today}

\begin{abstract}
We investigate the collapse kinetics of charged polymers (polyelectrolytes) induced by counterion condensation using coarse-grained molecular dynamics simulations. Under good solvent conditions, polyelectrolytes above the critical charge density ($A > A_c$) exhibit significantly faster collapse dynamics compared to neutral polymers, with dynamic scaling exponents ($\nu_c \sim 0.76$–$0.84$) distinctly smaller than those observed for neutral polymers ($\nu_c \approx 1.44$). This accelerated collapse is driven primarily by three mechanisms: (1) local charge neutralization due to counterion condensation, which facilitates immediate local compaction, (2) screening of long-range electrostatic repulsions, reducing the conformational search space, and (3) bridging interactions mediated by multivalent counterions, enhancing efficient formation of intra-chain contacts. We systematically explore the effects of polymer length, charge density, and counterion valency (monovalent, divalent, and trivalent) on collapse dynamics, demonstrating that increased counterion valency significantly lowers the critical charge density required for collapse and accelerates the collapse process. Our findings highlight the limitations of modeling charged biopolymers using purely neutral coarse-grained models, underscoring the importance of electrostatic interactions and counterion dynamics in determining their kinetic pathways. These insights may aid in better understanding the folding, organization, and dynamics of inherently charged biomolecules, such as proteins and nucleic acids.
\end{abstract}
\maketitle

\section{Introduction}

Many biological phenomena, such as protein folding and chromatin organization, involve charged polymers and their compaction~\cite{Wong-Rev,delaCruz2000,VirusRNA2006,VirusElectro2012,VirusGrossberg2016,woodson10,perlmutter2013viral,perlmutter2015mechanisms,zandi2020virus}. Understanding these biological functions requires not only knowledge of the final native structures but also information on the timescales of collapse and folding of such biopolymers, which is of paramount importance. The numerical values of equilibrium size, measured through the radius of gyration ($R_g$), and collapse times ($\tau_c$) depend on the detailed structure of the polymer and on external parameters such as temperature, co-solvent concentrations, external fields, ionic strengths, and solution pH. However, the scaling of $R_g$ and $\tau_c$ with polymer length $N$ is expected to be independent of specific structural details~\cite{dewey1993protein,hong2009scaling,rubinstein2003polymer}, as microscopic interactions are integrated into effective generic interactions at larger length and time scales. Coarse-grained models have thus been employed to study the large-scale structural and temporal properties of polymers and peptides, circumventing the computational cost of all-atom simulations~\cite{kmiecik2016coarse,noid2013perspective,gartner2019modeling,zhou2004polymer}. The study of polymer collapse using coarse-grained models remains an active area of research due to its connection with the early stages of protein folding~\cite{haran2012and,udgaonkar2013polypeptide}. The extended-to-collapsed transition in polymers can occur through various pathways depending on specific conditions and interactions. One well-known pathway involves changes in solvent quality, where poor solvent conditions with weaker polymer-solvent interactions promote polymer collapse; this can be induced by increasing the non-solvent concentration or altering the solvent's nature, manifesting as the well-known hydrophobic collapse of biopolymers in water~\cite{flory1953principles,de1979scaling,de1990polymers,clark2020water,zhou2004hydrophobic}. The addition of salts or other solutes to a polymer solution can also influence the transition by modifying the ionic strength or introducing specific interactions between polymer and solute molecules, which is particularly relevant for biopolymers like proteins and DNA~\cite{sherman2006coil,frerix2006exploitation}. Depletion interactions, where smaller particles or colloids are introduced as depletants into polymer solutions, drive the transition through the excluded volume effect, leading to polymer collapse~\cite{asakura1958interaction,asakura1954interaction,vrij1976polymers}. Crowding effects, particularly those involving attractive crowders, can also induce transitions due to specific interactions between polymers and solvent/crowder molecules~\cite{jiao2010attractive,kim2013crowding,antypov2008computer,heyda2013rationalizing,rodriguez2015mechanism,sagle2009investigating,huang2021chain,ryu2021bridging,brackley2020polymer,brackley2013nonspecific,barbieri2012complexity,garg2023conformational}. Systems of charged polymers in the presence of oppositely charged counterions, known as polyelectrolytes (PEs), serve as important model systems for understanding the conformational properties of highly charged polymers. Strongly charged PE chains collapse, regardless of solvent quality, when their charge density exceeds a critical threshold due to the condensation of oppositely charged counterions~\cite{brilliantov1998chain,liu2003polyelectrolyte,tom2016mechanism,varghese2011phase}.

The scaling of $R_g$ with $N$ is well established for both neutral and charged polymers: within Flory theory, $R_g \sim N^{3/5}$ in the extended regime, $R_g \sim N^{1/2}$ for theta-solvents, and $R_g \sim N^{1/3}$ in the collapsed regime~\cite{de1979scaling,rubinstein2003polymer}. In contrast, the collapse kinetics have been more extensively studied for neutral polymers using a variety of methods~\cite{de1985kinetics,grosberg1988role,byrne1995kinetics,kuznetsov1995kinetics,kuznetsov1996kinetic,pitard1998dynamics,pitard1999influence,klushin1998kinetics,halperin2000early,abrams2002collapse,chang2001solvent,kikuchi2005kinetics,lee2006mesoscopic,pham2008brownian,majumder2017kinetics,paciolla2020coarsening}. The dynamic exponent $\nu_c$, describing the scaling of $\tau_c$ with $N$ ($\tau_c \sim N^{\nu_c}$), is typically measured by monitoring $R_g$ or tracking the growth of cluster sizes along the collapse trajectory. The values of $\nu_c$ for neutral polymers, obtained from both simulations and theory, are summarized in Table~\ref{tab-0}. Unlike the well-established scaling exponents of $R_g$ with $N$, the values of $\nu_c$ show considerable variation. However, there is a general consensus that for neutral polymers, $\nu_c \approx 1.66$ in the absence of hydrodynamic interactions and $\nu_c \approx 1$ when hydrodynamic interactions are present.The value of $\nu_c$ for charged polymers remains less established. These polymers are often modeled as effective neutral polymers, both to simplify calculations and because additional salt screening converts the system into one with effective short-range interactions~\cite{kang2015effects,moreno2016concentrated,qin2013effects,banks2018intrinsically,forrey2006langevin,brahmachari2022shaping}. As a result, it is implicitly assumed that the collapse timescales of neutral and charged polymers are similar. However, since most biopolymers, such as proteins and DNA, are inherently charged, establishing the collapse kinetics of charged polymers is critical for accurately modeling their behavior.

Recently, we investigated the collapse kinetics of polyelectrolytes (PEs) in poor solvents near the critical charge density, the threshold above which counterion condensation neutralizes the polymer's effective charge and induces collapse, and demonstrated that PEs exhibit different collapse kinetics compared to neutral polymers, as indicated by a smaller collapse scaling exponent, $\nu_c$\cite{ghosh2021kinetics}. However, to make a direct comparison with the values of $\nu_c$ summarized in Table\ref{tab-0}, it is necessary to measure $\nu_c$ at much higher charge densities, as the kinetics near the critical point may differ significantly. A recent study~\cite{yuan2024collapse} examined the collapse dynamics of charged polymers under hydrodynamic interactions and emphasized the significant impact of electrostatic interactions on the scaling behavior $\tau_{c} \sim N^{\nu_{c}}$. The study reported an increased value of $\nu_c \approx 1.4$ for charged polymers, compared to $\nu_c \approx 0.94$ for neutral polymers. It is important to note that this collapse transition was induced by quenching from good solvent conditions to poor solvent conditions for weakly charged polymers without explicitly considering whether the polymer charge density exceeded the critical charge density ($A_c$). Below this critical threshold, polymers remain inherently repulsive, influencing collapse kinetics differently compared to systems above $A_c$, where counterion condensation neutralizes charges. The influence of the charge density of the PE chain and counterion valency on the counterion-induced collapse transition, however, remains unexplored. 

\begin{table}
\caption{The known values of $\nu_c$ [collapse time $\tau_c \sim N^{\nu_c}$]  for neutral polymers obtained from different methods. HI (no HI) stands for results in presence (absence) of hydrodynamics) \label{tab-0}}
\begin{ruledtabular}
\begin{tabular}{lll}
Reference & $\nu_c$ (HI) & $\nu_c$ (no HI)  \\
\hline
de Gennes (1985)~\cite{de1985kinetics} &- & 2.0  \\
Grosberg $et~al. (1988)~$\cite{grosberg1988role} &- &2.0\\
Kuznetsov$et~al.$ (1995)~\cite{kuznetsov1995kinetics}&- &2.0 \\
Kuznetsov$et~al.$ (1996)~\cite{kuznetsov1996kinetic}&1.0&1.67 \\
Byrne $et~al.$ (1995)~\cite{byrne1995kinetics}& -&1.67\\
Buguin $et~al.$ (1996)~\cite{buguin1996collapse}  & 1.0 & 2.0 \\
Halperin $et~al.$ (2000)~\cite{halperin2000early} & - & $1.2$\\
Klushin $et~al.$ (1998)~\cite{klushin1998kinetics} & 0.93 & 1.60 \\
Pitard $et~al.$ (1998)~\cite{pitard1998dynamics} & 1.0 & 1.66 \\
Pitard (1999)~\cite{pitard1999influence}& 1.0 & 1.66\\
Chang $et~al.$ (2001)~\cite{chang2001solvent} & 0.57 &1.50\\
Abrams $et~al.$ (2002)~\cite{abrams2002collapse} & 0.83 & 1.5\\
Kikuchi $et~al.$ (2005)~\cite{kikuchi2005kinetics}&  1.33 & 2.0  \\
Lee $et~al.$ (2006)~\cite{lee2006mesoscopic}& 1.0 &-\\
Pham $et ~al.$ (2008)~\cite{pham2008brownian} &1.01 & 1.35  \\
Guo $et al.$ (2011)~\cite{guo2011coil} & 1.0 & -\\
Majumder $et~al.$ (2017)~\cite{majumder2017kinetics} & - &1.8 \\
Christiansen $et~al.$ (2017)~\cite{christiansen2017coarsening} & - &1.61 \\
Majumder $et~al.$ (2018)~\cite{majumder2018scaling} & - &1.61, 1.0 \\
Majumder $et~al.$ (2019)~\cite{majumder2019pearl} & 0.5&- \\
Majumder $et~al.$ (2020)~\cite{majumder2020understanding} & - &1.8, 1.0, 1.61 \\
Schneider $et~al.$ (2020)~\cite{schneider2020different} & 0.94 & -\\
Paciolla $et~al.$ (2020)~\cite{paciolla2020coarsening}& -&1.67\\
Chauhan $et~al.$ (2022)~\cite{chauhan2022delayed}&1.00 &- \\
Yuan $et~al.$ (2024)~\cite{yuan2024collapse}&0.94 &- \\
This work&-&1.445
	\end{tabular}
	\end{ruledtabular}
\end{table}

In this paper, using extensive molecular dynamics (MD) simulations in the absence of hydrodynamic interactions, we study the collapse kinetics of polyelectrolytes (PEs) at charge densities significantly higher than the critical charge density, under good solvent conditions, with counterion condensation~\cite{brilliantov1998chain,liu2003polyelectrolyte,sarragucca2003structure,wei2007role,wei2010effect,tom2016mechanism,varghese2011phase} as the central mechanism of collapse. We begin by establishing the collapse kinetics of a neutral polymer as a reference. Studying PE collapse kinetics in good solvent conditions, at charge densities significantly above the critical threshold for counterion condensation ($A > A_c$), offers the advantage that the collapse is driven solely by electrostatic interactions and counterion dynamics, without confounding effects of solvent quality changes. Our main result shows that, in the absence of additional salt, the dynamic exponent for PEs is $\nu_c \sim 0.76$–$0.84$. These values are markedly different from the well-established $\nu_c \approx 1.44$ for neutral polymers in the absence of hydrodynamic interactions (see Table~\ref{tab-0}). Additionally, the observation that $\nu_c$ is less than one suggests that collapse process for charged polymers is faster than ballistic collapse.

\section{Method}
We model the flexible PE chain as a coarse grained linear bead spring model consisting of  $N$ monomers of charge $-q$ (measured in units of $e$) linked together by harmonic springs with  bond interaction energy
\begin{equation}
U_{bond}(r)=\frac{1}{2} k(r-r_0)^2,
\end{equation}
where $k$ is the spring constant, $r_0$ is the equilibrium bond length and $r$ is the distance between the bonded monomers. We set $r_0=1.12 \sigma$, $k=500 \epsilon \sigma^{-2}$, where $\sigma$ and $\epsilon$ are  parameters of the Lennard Jones (LJ) potential discussed below. The $N/Z$ neutralizing counterions are particles with charge $+Z q$, where $Z$ is the valency.

All the non-bonded pairs of particles interact through excluded volume interactions as well as long ranged electrostatic interactions. The excluded volume interactions are purely repulsive and modeled by the shifted LJ potential
\begin{equation}
\begin{aligned}
U_{\mathrm{LJ}}(r)&=4\epsilon\left[\left(\frac{\sigma}{r}\right)^{12}-\left(\frac{\sigma}{r}\right)^{6} \right], 
\end{aligned}
\label{eq.2}
\end{equation}
where the interaction is cutoff at separation $r_c=2^{1/6} \sigma$. The van der Waals radius, $\sigma$ and the interaction parameter $\epsilon$ are the same for all pairs of particles.

The PE,  counterions and salt  are assumed to be
in a medium of uniform relative dielectric constant $\varepsilon$. The electrostatic energy between two charges $q_{1}$ and $q_{2}$ separated by distance $r$ is given by
\begin{equation}
U_{C}(r)= \frac{q_1 q_2}{4\pi\varepsilon\varepsilon_{0} r},
\label{eq.1}
\end{equation}
where $\varepsilon_{0}$ is the dielectric permittivity of vacuum.

The relative strength of electrostatic interactions among charged monomers along the PE chain is characterized by a dimensionless parameter $A$:
\begin{equation}
A=\frac{q^{2}\ell_B}{r_0},\quad \ell_B=\frac{e^2}{4\pi\varepsilon\varepsilon_{0} k_{B}T},
\label{eq.4}
\end{equation}
where $\ell_B$ is the Bjerrum length~\cite{khokhlov1994}  at which the electrostatic energy between two elementary charges $e$ is comparable  to the thermal energy $k_{B}T$, 
where $k_B$ is  the Boltzmann constant and $T$ is the temperature. The strength of electrostatic interactions, $A$, can be tuned by either  changing $q$ while keeping  $\varepsilon$ fixed or by changing $\varepsilon$ while keeping $q$ fixed. In our simulations, we systematically scale the monomer charges $q$ to vary $A$.
Lengths, temperatures and times are
given in units of $\sigma$, $\epsilon /k_{B}$, and $
\sqrt{m\sigma^{2}/\epsilon}$ respectively. 

All the simulations are carried out by molecular dynamics simulation package LAMMPS~\cite{LAMMPS} in NVT ensemble with periodic boundary conditions. The equations of motion are integrated via a time step of 0.001 and the data are saved in every 5000 steps. The box size $L$ is chosen large enough to guarantee that the PE chain in good solvent condition does not interact with its own periodic images. For different $N$,   $L$ is chosen such that the monomer density is fixed at $N/L^{3} = 0.895 \times 10^{-6} \sigma^{-3}$.  A Langevin thermostat is  utilized to examine the kinetics of PE collapse. The long-range Coulomb interactions are evaluated using the particle-particle/particle-mesh (PPPM) technique. 

For simulations of the collapse dynamics of the PE chain, the initial configuration at $A=0.2$ is generated by performing a self-avoiding random walk on a simple cubic lattice, along with counterions. This configuration is then subjected to a long equilibrium run of $10^{7}$ MD steps to ensure a uniform distribution of counterions throughout the simulation box. From this equilibrium run, 50 statistically independent initial configurations of PE-counterions in good solvent conditions at $A=0.2$ are selected, and each system is further equilibrated for a sufficiently long time. During these equilibration steps, a constant temperature is maintained using a Nose-Hoover thermostat to facilitate the system’s relaxation to equilibrium. Each configuration is then instantaneously quenched to the desired charge density in good solvent conditions. For the final simulations of collapse dynamics, a Langevin thermostat is employed, as described earlier. We simulate multiple systems with varying charge densities ($A$), counterion valencies ($Z$), and polymer lengths ($N$). A summary of the equilibrium simulations performed to determine the critical charge density for collapse is provided in Table~\ref{tab-1}, while the details of the simulations used to study collapse dynamics are presented in Table~\ref{tab-2}.
\begin{table}		
	\caption{The range of values of valency ($Z$), charge density ($A$) and number of monomers ($N$)  in the MD simulations to determine $R_g$ in equilibrium.\label{tab-1}}
	\begin{ruledtabular}
		\begin{tabular}{lll}
$N$ & $Z$  & $A$  \\
\hline
402 & 1 &  0.5-13 \\
402 & 2 & 0.5-7 \\
402 & 3 & 0.5-4 \\
204 & 1  &  0.5-12\\
204 & 2 &  0.5-7 \\	
204 & 3 &  0.5-4 \\
		\end{tabular}
	\end{ruledtabular}
\end{table}
\begin{table}		
\caption{The range of values of valency ($Z$), charge density ($A$), number of monomers ($N$),  and histories ($H$) in the MD simulations to determine variation of $R_g$ with time.  \label{tab-2}}
\begin{ruledtabular}
\begin{tabular}{llcc}
$Z$ & $A$ & $N$ & $H$\\
\hline
1 & 20.0 & \makecell{$384$, $288$, 240, $192$\\144, $96$, 72, $48$} & 50 \\
\hline
2 & 8.0 & \thead{$384$, $288$, 240, $192$\\144, $96$, 72, $48$} & 50 \\
\hline
3 & 5.0 & \thead{$384$, 336, $288$, 240\\$192$, 144, $96$, 72, $48$} & 50\\	
\end{tabular}
\end{ruledtabular}
\end{table}
\\

To validate our results for the polyelectrolyte system, we also simulate neutral polymer systems with different polymer lengths ($N=60, 120, 180, 240$). The typical simulation protocol for the neutral polymer system is as follows: The initial configuration of a neutral polymer chain is randomly generated within a large simulation box, and the same bead-spring model described in Eq.~\ref{eq.1} is employed. The dimensions of the box are chosen to match the particle density of the polyelectrolyte system. Molecular dynamics simulations are performed using a time step of 0.001 with the interaction potential of a good solvent. Once equilibration is achieved, the equilibrium properties are averaged over the last $10^{6}$ MD steps of a $10^{7}$ step-long simulation, with sampling every 1000 MD steps. For collapse dynamics simulations, 50 statistically independent configurations are selected from the good solvent condition by using the closest available final configuration from a previous simulation. Each system is equilibrated by assigning initial random velocities using a random number generator with a specified seed at the desired temperature. Depending on the degree of polymerization, approximately $1.4 \times 10^{6}$ time steps are typically required to generate each statistically independent configuration. Subsequently, each chain undergoes quenching by changing the nature of the solvent from good to poor. In good solvent conditions, the monomer-monomer interaction is given by a shifted Lennard-Jones (LJ) potential or Weeks-Chandler-Andersen (WCA) potential as described in Eq.~\ref{eq.2}. In poor solvent conditions, a Lennard-Jones potential with a larger cutoff ($r_{c}=2.5\sigma$) is employed to include an attractive interaction between all beads, in addition to the excluded volume interaction.

\section{Results}
The collapse of both neutral polymers and polyelectrolytes (PEs) is identified by monitoring the radius of gyration, $R_g$, defined as:
\be
R_g^2 =\frac{1}{N} \sum_{i=1}^N \left| {\bf r}_i - {\bf R}_{CM} \right |^2,
\ee
where ${\bf r}i$ is the position of the $i$th monomer and ${\bf R}{CM}$ is the polymer’s center of mass. The collapse trajectories (averaged over 50 statistically independent initial configurations) are computed for neutral and charged polymers of varying chain lengths $N$. We define $\tau_{peak}$ as the time when the polymer reaches its maximum initial extension, after which the polymer begins to collapse. The mechanisms driving collapse differ between neutral polymers and PEs. For neutral polymers, collapse occurs due to an instantaneous quench from good solvent to poor solvent conditions. In contrast, the collapse of PEs is driven by counterion condensation, without any change in solvent quality. The simulated PEs lie in the charge density regime $A > A_c$, where $A_c$ is the critical charge density determined by Manning condensation.

\begin{figure}
\centering
\includegraphics[width=0.9\columnwidth]{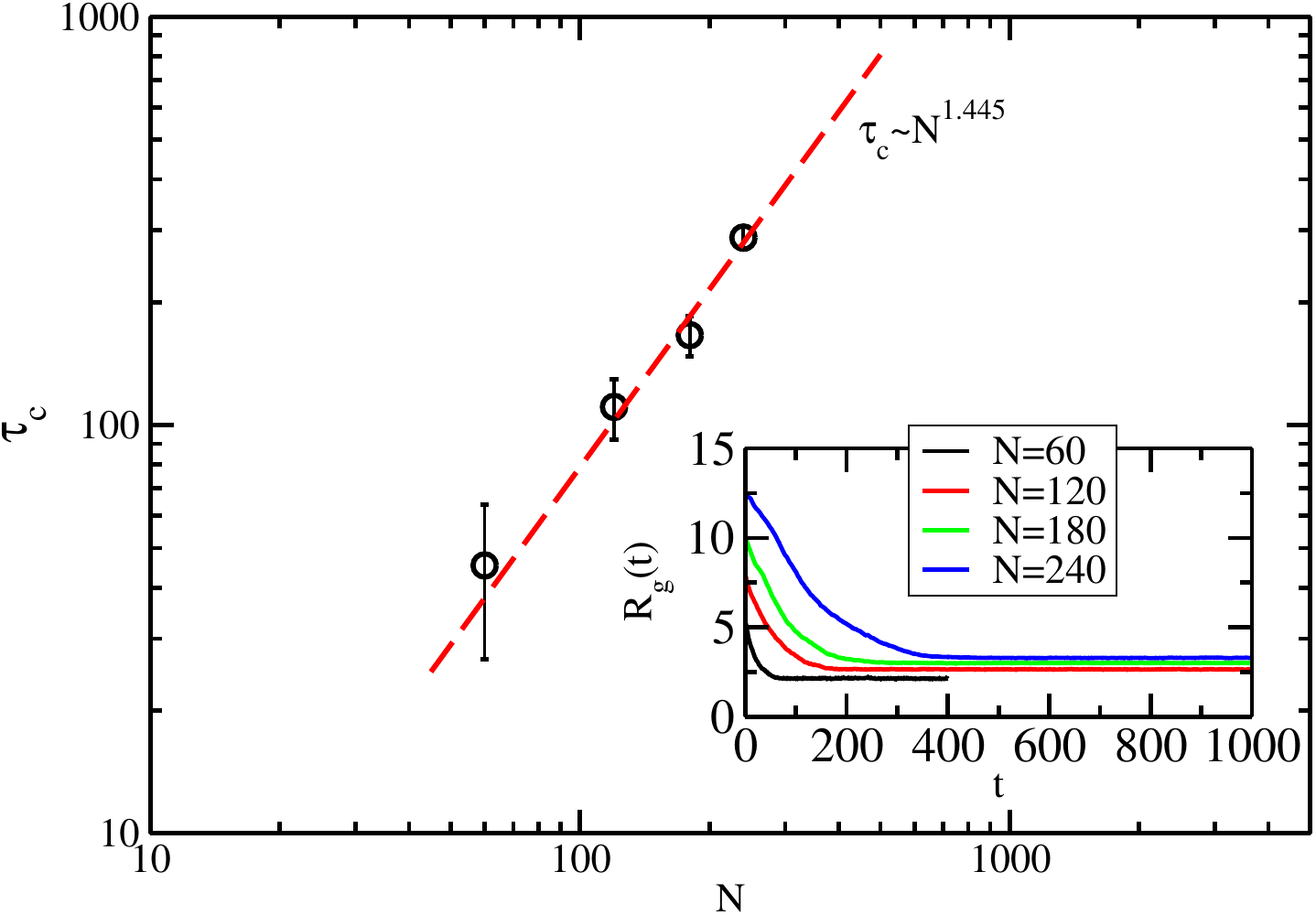}
\caption{Variation of the collapse time, $\tau_c$, with polymer length, $N$, for neutral polymers. The inset shows representative collapse trajectories of the radius of gyration, $R_g(t)$, as a function of time for different chain lengths ($N=60, 120, 180, 240$). Collapse times ($\tau_c$) are obtained from averaged trajectories (over 50 statistically independent initial configurations) and error bars represent the standard deviation across these independent simulations.}
\label{fig1}   
\end{figure}

The collapse time, $\tau_c$, is precisely defined as the time when the polymer radius reaches 99\% of its equilibrium value, measured from $\tau_{peak}$~\cite{schneider2020different,pham2008brownian}, according to:
\begin{equation}
R_g(\tau_c+\tau_{peak} ) =R_g^{eq}+0.01\left[R_g(\tau_{peak})-R_g^{eq}\right].
\end{equation}
$\tau_c$ is measured for each trajectory and averaged over 50 trajectories and is repeated for each $N$ value. The collapse exponent $\nu_c$ is then extracted from $\tau_c$ dependence on polymer length, $N$. Following this procedure, we obtain the scaling exponent $\nu_c$ from the relationship $\tau_c \sim N^{\nu_c}$ for neutral polymers, as shown in Fig.\ref{fig1}. We find $\nu_c=1.445$, consistent with previous theoretical and simulation results reported in the literature (see Table\ref{tab-0})~\cite{klushin1998kinetics,pitard1998dynamics,pitard1999influence,chang2001solvent,abrams2002collapse,kikuchi2005kinetics,pham2008brownian}. 

\begin{figure}
\centering
\includegraphics[width=0.6\columnwidth]{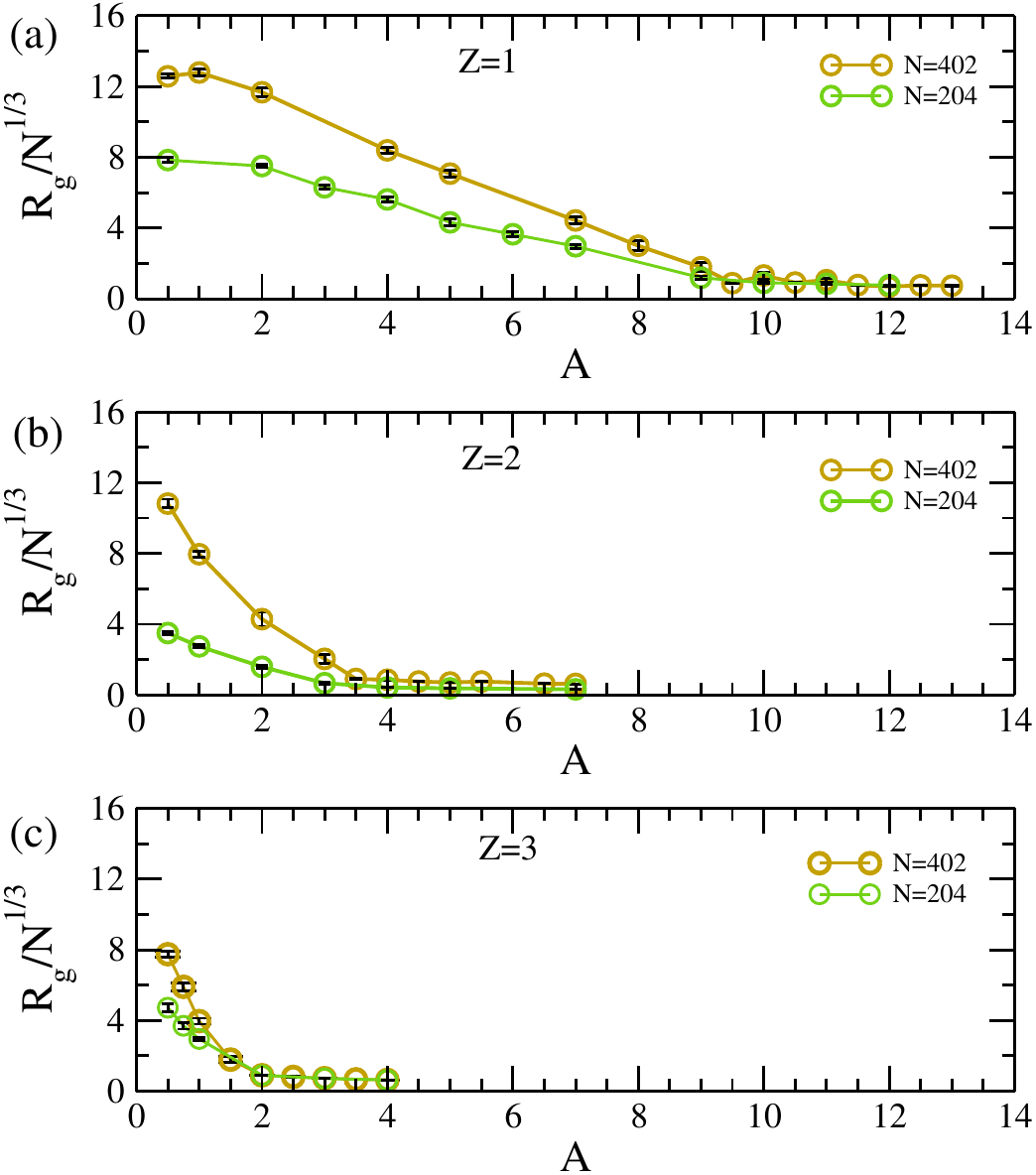}
\caption{The scaled equilibrium radius of gyration, $R_g^{eq}/N^{1/3}$, as a function of the charge density, $A$, for polyelectrolyte chains of lengths $N=204$ and $N=402$, interacting with (a) monovalent, (b) divalent, and (c) trivalent counterions. The critical charge density, $A_c$, is identified as the charge density at which $R_g^{eq}/N^{1/3}$ becomes independent of the polymer length, indicating the onset of collapse driven by counterion condensation. Error bars represent the standard deviation from averaging over multiple independent simulations.}
\label{fig2}
\end{figure}

For polyelectrolyte (PE) systems, we first determine the critical charge density, $A_c$, which corresponds to the threshold beyond which counterion condensation neutralizes electrostatic repulsions, triggering the polymer collapse transition under good solvent conditions. Initially, an extended PE configuration is equilibrated at a low charge density ($A = 0.2$), after which we incrementally increase the charge density to values exceeding the critical threshold. The equilibrium radius of gyration, $R_g^{eq}$, is calculated as an average over the final $10^{6}$ steps of a $10^{7}$-step equilibrium simulation. Figure~\ref{fig2} shows the variation of the scaled equilibrium radius of gyration, $R_g^{eq}/N^{1/3}$, with the charge density $A$ for two polymer chain lengths ($N = 204$ and $N = 402$) and for monovalent ($Z=1$), divalent ($Z=2$), and trivalent ($Z=3$) counterions. At low charge densities, the polymer remains in an extended state, and $R_g^{eq}/N^{1/3}$ shows clear dependence on chain length $N$. However, as $A$ increases, $R_g^{eq}/N^{1/3}$ monotonically decreases and eventually becomes independent of $N$, indicating a transition into a collapsed regime. We identify the critical charge density, $A_c$, as the value of $A$ above which the scaled equilibrium radius of gyration no longer depends significantly on polymer length. The observed critical charge densities decrease with increasing counterion valency, yielding $A_c \approx 9.5$, $4.0$, and $2.0$ for $Z=1$, $2$, and $3$, respectively, consistent with the expectation that higher-valency counterions condense more efficiently.

\begin{figure}
\centering
\includegraphics[width=\columnwidth]{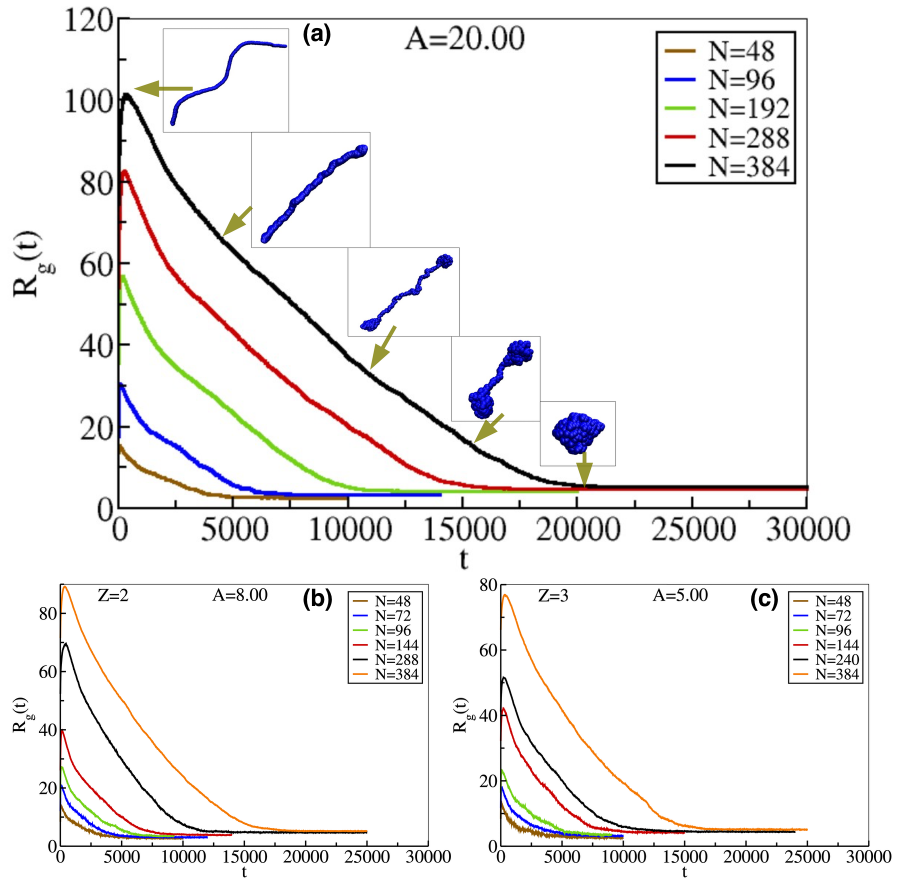}
\caption{The variation of the history averaged radius of gyration $R_g(t)$ with time $t$ for different polymer lengths $N$. The data are for (a) $Z=1$ and $A=20.00$, (b) $Z=2$ and $A=8.00,$ and (c) $Z=3$ and $A=5.00$. The snapshots correspond to the different times indicated in the time evolution for $Z=1$ and $N=384$. }
\label{fig3}
\end{figure}

After identifying the critical charge density $A_c$ for different counterions, we quench the system from an extended phase ($A=0.2$) to $A=20.0$, $8.0$, and $5.0$ for $Z=1$, $2$, and $3$, respectively, to study the collapse kinetics of PEs induced by counterion condensation. It is important to note that the solvent condition remains unchanged and is retained as a good solvent throughout the simulations. For charged PEs, the charge density along the polymer chain influences the transition from an extended state (at low charge density) to a collapsed state (at high charge density), analogous to the solvent-induced collapse observed in neutral polymers. In Fig.\ref{fig3}, we present the history-averaged collapse trajectories (averaged over 50 simulations) of PEs in the presence of counterions with $Z=1$, $2$, and $3$ for different values of $N$. For all three counterion valencies considered, $R_g$ increases from its initial value until the time $\tau_{peak}$, after which it decreases to its equilibrium value as the PE collapses (see snapshots in Fig.\ref{fig3}(a) for $Z=1$). The collapse time, $\tau_{c}$, defined as the time when $R_g \approx R_g^{eq}$, is observed to increase with system size, similar to the case of neutral polymers. We define $\tau_{c}$ as the time required for the PE to complete 99\% of its collapse, measured from $\tau_{peak}$, following the same procedure used for neutral polymers and as described in Refs.~\cite{schneider2020different, pham2008brownian}. The collapse time $\tau_{c}$ is measured for each trajectory and then averaged over 50 independent trajectories.

\begin{figure}
\centering
\includegraphics[width=\columnwidth]{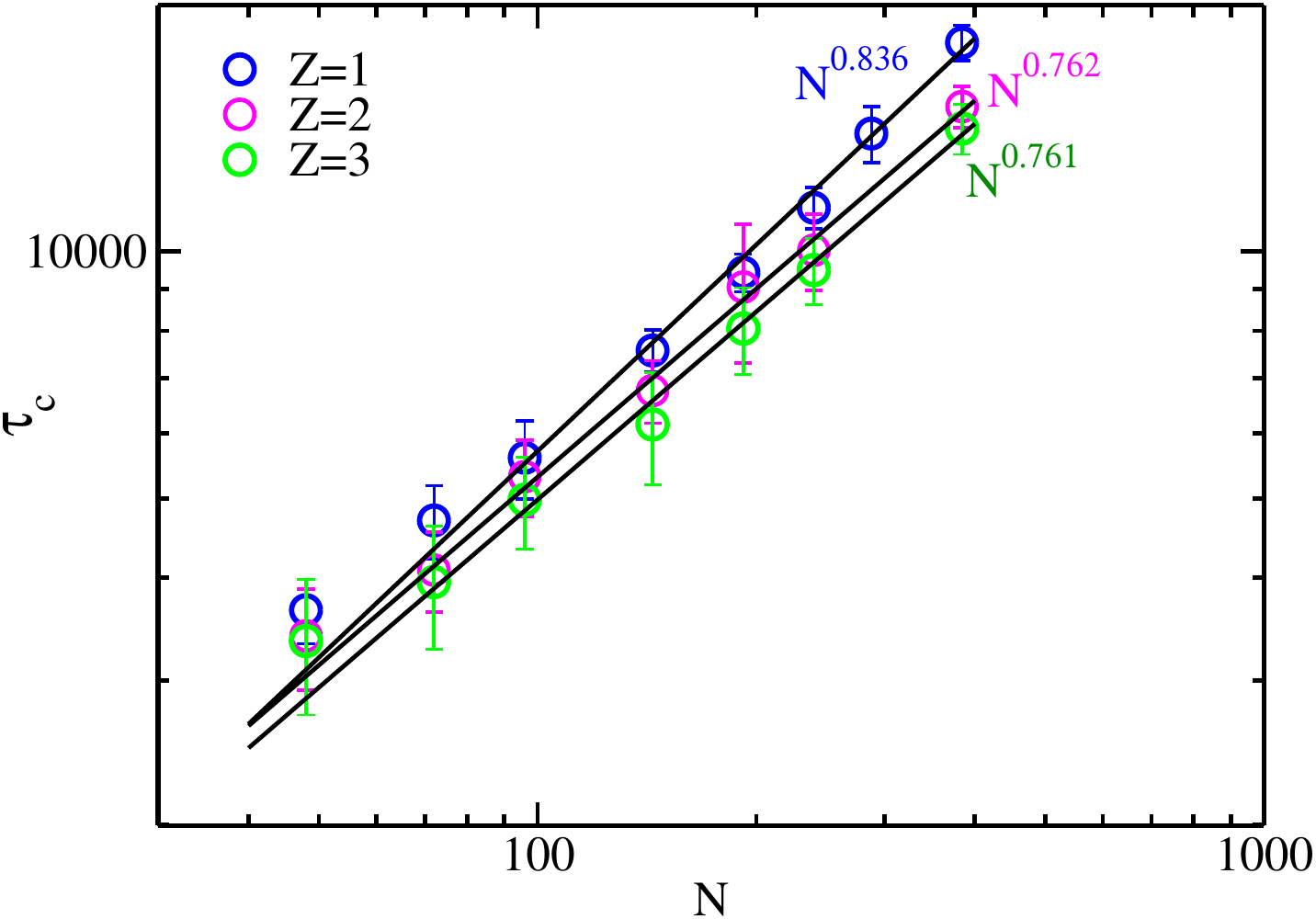}
\caption{The variation of $\tau_c$ with $N$ for monovalent, divalent and trivalent counterions.}
\label{fig4}
\end{figure}

After identifying the critical charge density, $A_c$, for different counterion valencies, we investigate the collapse kinetics of PEs by quenching from an initially extended configuration ($A=0.2$) to charge densities significantly above the critical threshold ($A=20.0$, $8.0$, and $5.0$ for $Z=1$, $2$, and $3$, respectively). It is important to emphasize that the simulations are performed under good solvent conditions throughout, isolating collapse induced solely by counterion condensation rather than solvent quality changes. Figure~\ref{fig3} shows the history-averaged radius of gyration, $R_g(t)$, for different polymer lengths ($N$) and counterion valencies. For all three valencies, the radius of gyration initially increases, reaching a maximum at time $\tau_{peak}$, before subsequently decreasing to its equilibrium collapsed size. Representative polymer conformations at various stages of collapse for $Z=1$ and $N=384$ are shown as snapshots in Fig.~\ref{fig3}(a). The collapse time $\tau_c$ is quantitatively defined as the time required for the polymer radius to reach 99\% of its equilibrium value, measured from $\tau_{peak}$, following the same methodology used previously for neutral polymers~\cite{schneider2020different, pham2008brownian}. Values of $\tau_c$ are obtained by averaging over 50 independent collapse trajectories for each polymer length and valency studied.

In Fig.~\ref{fig4}, we present the variation of the collapse time $\tau_c$ with polymer length $N$ for monovalent ($Z=1$), divalent ($Z=2$), and trivalent ($Z=3$) counterions, clearly displaying a power-law scaling behavior: $\tau_c \sim N^{\nu_c}$. The fitted dynamic scaling exponents are $\nu_c = 0.836$, $0.762$, and $0.761$ for monovalent, divalent, and trivalent counterions, respectively. These exponents are significantly lower than the value ($\nu_c \approx 1.445$) obtained for neutral polymers in the absence of hydrodynamic interactions, indicating substantially faster collapse kinetics for PEs. The scaling exponent for monovalent counterions ($\nu_c = 0.836$) suggests a collapse dynamics faster than diffusive but slower than ballistic transport. In contrast, the smaller exponents obtained with divalent ($\nu_c = 0.762$) and trivalent ($\nu_c = 0.761$) counterions indicate even more accelerated, super-ballistic collapse kinetics. The increased rate of collapse observed for higher-valency counterions is attributable to more efficient local charge neutralization and counterion bridging, which reduce electrostatic repulsions and expedite intra-chain contact formation. Moreover, the close similarity in scaling exponents between divalent and trivalent counterions suggests a saturation effect: beyond a certain valency threshold, further increasing counterion valency does not substantially enhance the collapse rate. The linearity and consistency of the scaling data on a log-log scale for all counterion valencies and polymer lengths underscore the robustness of our results. The observed behavior clearly demonstrates that electrostatic interactions, mediated through counterion condensation, significantly dominate and accelerate the collapse dynamics of charged polymers compared to neutral polymers.

\begin{figure}
\centering
\includegraphics[width=1.0\columnwidth]{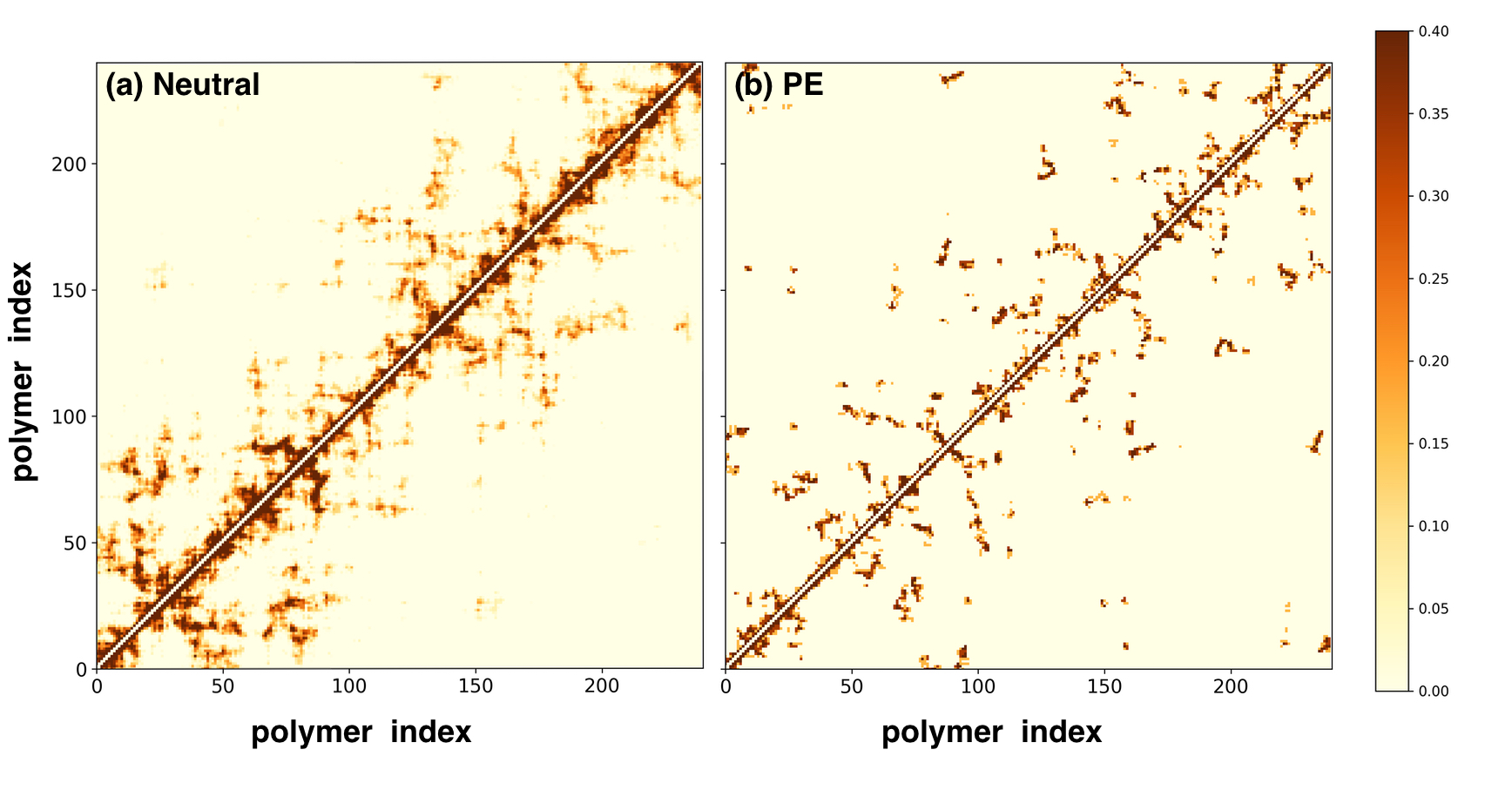}
\caption{Contact probability maps illustrating the spatial organization of intra-chain contacts during the collapse of (a) neutral polymers and (b) polyelectrolytes with monovalent counterions }
\label{fig5}
\end{figure}

\begin{figure}
\centering
\includegraphics[width=1.0\columnwidth]{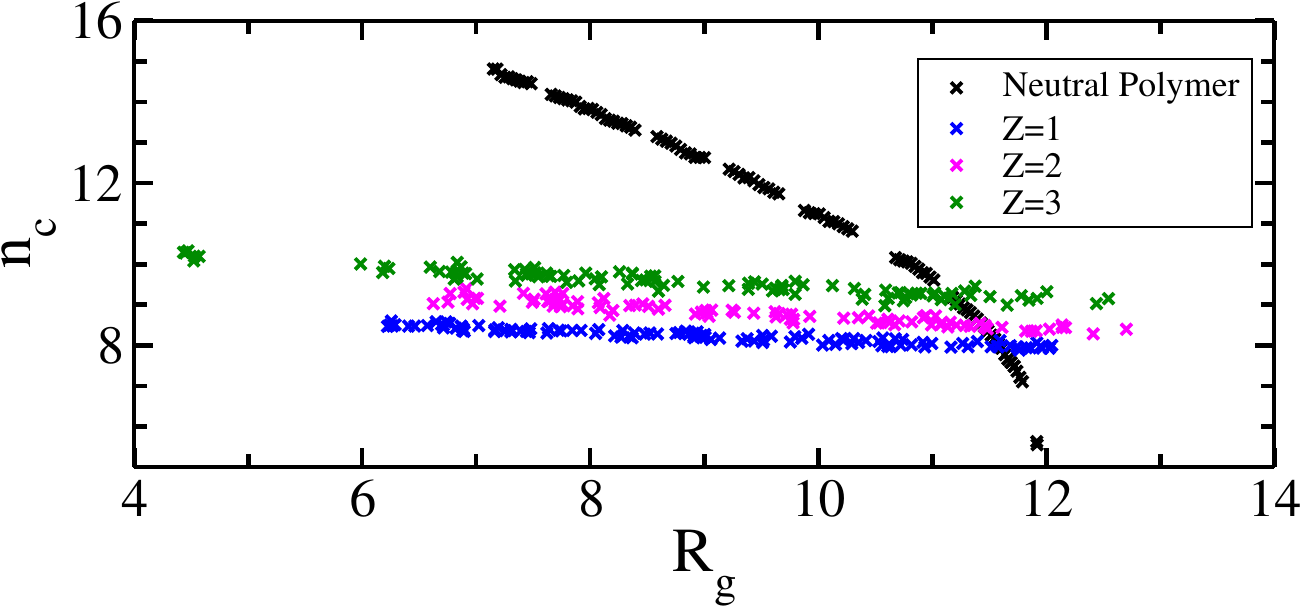}
\caption{Average number of intra-chain contacts, $n_c$, plotted as a function of radius of gyration $R_g$ for neutral polymers and polyelectrolytes with different counterion valencies.  }
\label{fig6}
\end{figure}

We carefully analyze the mechanisms underlying the differences in collapse dynamics between neutral polymers and polyelectrolytes (PE), guided by Figures~\ref{fig5} and ~\ref{fig6}. Figure5 displays contact probability maps for neutral and charged polymers during collapse, offering visual insight into intra-chain interactions. For neutral polymers (Fig.~\ref{fig5}a), the contact map exhibits prominent diagonal and substantial off-diagonal features, indicating extensive local and long-range contacts. This pattern suggests a predominantly diffusion-driven collapse mechanism, where polymer segments extensively explore conformational space, gradually forming contacts diffusely. Such a mechanism inherently exhibits a stronger dependence on chain length $N$, contributing to the (Fig.~\ref{fig5}b) exhibit sharper diagonal contacts, highlighting strongly localized intra-chain interactions. A closer examination reveals that, although localized interactions dominate, weaker but discernible off-diagonal contacts are present. These weaker off-diagonal elements indicate that long-range contacts, while significantly fewer compared to neutral polymers, do occur during PE collapse. These contacts likely arise from bridging interactions facilitated by condensed multivalent counterions, efficiently linking distant polymer segments. As counterion valency $Z$ increases from monovalent to divalent and trivalent, diagonal contacts become progressively sharper, reinforcing that multivalent ions predominantly intensify local compaction but also facilitate occasional bridging interactions.

Collapse dynamics in polyelectrolytes thus differ fundamentally from neutral polymers. For neutral polymers, collapse kinetics are dominated by hydrophobic (volume-driven) interactions, involving substantial conformational exploration. This diffuse and stochastic search results in pronounced size dependence and larger scaling exponents ($\nu_c > 1$). Conversely, polyelectrolyte collapse kinetics are significantly influenced by localized electrostatic interactions and counterion-mediated bridging. Immediate local charge neutralization substantially reduces repulsive interactions, while bridging by multivalent counterions facilitates efficient local compaction and occasional long-range contacts, significantly reducing the chain-length dependence and thus yielding lower scaling exponents ($\nu_c < 1$). These interpretations drawn from Fig.~\ref{fig5} are explicitly supported and clarified by Fig.~\ref{fig6}. Figure~\ref{fig6} plots the average number of intra-chain contacts against the radius of gyration ($R_g$), explicitly illustrating structural and mechanistic nuances during polymer collapse. In the intermediate-size regime ($R_g \gtrsim 10$), the number of intra-chain contacts systematically increases with counterion valency ($Z$). Higher-valency counterions (divalent and trivalent) consistently produce more intra-chain contacts than neutral polymers at intermediate polymer sizes. This occurs because higher-valency counterions strongly facilitate numerous bridging interactions between monomers separated widely along the polymer chain, rapidly compacting distant segments. This bridging contrasts with the stochastic exploration typical of neutral polymers, significantly reducing necessary conformational exploration and greatly accelerating collapse. Thus, the lower scaling exponent for charged polymers directly correlates with these enhanced bridging interactions. In the late-stage collapse regime ($R_g < 10$), Fig.~\ref{fig6} demonstrates that neutral polymers achieve a higher number of contacts compared to charged polymers, irrespective of valency $Z$. This indicates that neutral polymers ultimately attain denser conformations during final collapse stages compared to polyelectrolytes. This denser packing in neutral polymers suggests that volume-driven collapse allows tighter packing without residual electrostatic repulsions or steric hindrance from condensed ions. In contrast, polyelectrolytes, despite rapid initial collapse, encounter constraints from residual counterions, limiting dense equilibrium packing. This is primarily due to the presence of condensed counterions inside the collapse phase of charged polymers, which is not the case for neutral polymers in the collapsed phase.

Summarizing these observations and analyses, we propose a refined explanation for differences in dynamic scaling exponents between neutral and charged polymer collapses, explicitly illustrated in Figures~\ref{fig5} and \ref{fig6}. Neutral polymers experience slower, hydrophobic-driven collapse, necessitating extensive conformational space exploration and resulting in stronger size dependence ($\nu_c > 1$). Polyelectrolytes, by contrast, rapidly undergo initial local collapse via electrostatic neutralization significantly enhanced by multivalent counterion bridging interactions. This bridging sharply elevates intra-chain contacts at intermediate collapsed sizes, drastically reducing the required conformational search and lowering the scaling exponent ($\nu_c < 1$). Nonetheless, neutral polymers ultimately achieve more densely packed equilibrium conformations, highlighting a nuanced interplay between collapse kinetics and equilibrium structural outcomes.

\section{Discussion and Conclusion}
The literature on the collapse kinetics of charged polymers remains sparse, despite their prevalence in biological systems. Understanding the collapse and subsequent folding or organization of charged biopolymers, such as proteins and nucleic acids, is crucial for elucidating their biological functions. Although the collapse of neutral polymers has been extensively studied, the dynamics of charged polymers, or polyelectrolytes (PEs), remain relatively poorly understood, particularly concerning the role of electrostatic interactions and counterion condensation. Recent experimental studies underscore the importance of electrostatics in the early stages of biopolymer folding. For instance, time-resolved solid-state NMR experiments have demonstrated rapid initial collapse of proteins, followed by prolonged rearrangement to reach their final folded structures\cite{wilson2024experimental}. This experimental finding aligns with our earlier simulation results\cite{ghosh2021kinetics}, where charged polymers exhibited faster collapse kinetics compared to neutral polymers, but slower rearrangement due to fluctuations in counterion binding and charge distribution.

In this study, we investigated the collapse kinetics of charged polymers under good solvent conditions, focusing solely on electrostatic interactions and counterion condensation. Our extensive molecular dynamics simulations confirm that polyelectrolytes collapse more rapidly than neutral polymers, even under good solvent conditions. This result, consistent with our prior observations in poor solvent conditions~\cite{ghosh2021kinetics}, suggests that the faster collapse in polyelectrolytes primarily arises from electrostatic interactions and counterion condensation, independent of solvent quality. To quantitatively assess collapse kinetics, we defined the collapse time $\tau_c$ as the duration required for the polymer radius of gyration $R_g$ to transition from its maximum to its equilibrium value. The scaling of $\tau_c$ with polymer length $N$ follows the power-law relation, $\tau_c \sim N^{\nu_c}$, where $\nu_c$ denotes the dynamic scaling exponent.  Remarkably, we observe significantly lower $\nu_c$ values for polyelectrolytes (0.66–0.84) compared to neutral polymers ($\nu_c \approx 1.6$, Table~\ref{tab-0}). This substantial difference highlights distinct collapse mechanisms between neutral and charged polymers.

Neutral polymers collapse through a diffusion-limited process characterized by random formation of both local and long-range contacts. Conversely, polyelectrolytes undergo a more localized, directed collapse driven by counterion condensation and electrostatic interactions. Contact maps in Fig.~\ref{fig5} illustrate these differences clearly. Neutral polymer contact maps (Fig.~\ref{fig5}a) display a diffuse diagonal line alongside substantial off-diagonal contacts, consistent with extensive conformational space exploration and stochastic search for intra-chain interactions. Polyelectrolyte contact maps (Fig.~\ref{fig5}b) exhibit sharper diagonal features, reflecting dominant local interactions facilitated by charge neutralization from condensed counterions. The fewer off-diagonal contacts observed for PEs support the dominance of localized compaction during collapse. An intriguing finding from our results is the observation of dynamic scaling exponents $\nu_c < 1$, indicating super-ballistic collapse in polyelectrolytes. Classically, diffusion-driven collapse leads to $\nu_c > 1$, reflecting slow conformational exploration. However, the $\nu_c < 1$ observed here implies faster-than-ballistic dynamics, likely driven by persistent intra-chain contacts and high counterion mobility along the polymer backbone, as predicted by counterion fluctuation theory~\cite{brilliantov1998chain}. These counterion-driven bridging events significantly accelerate local dynamics, further enhancing collapse speed. Thus, the collapse kinetics of charged polymers fundamentally differ from those of neutral polymers, driven predominantly by electrostatic interactions and counterion condensation. The rapid initial collapse results from localized charge neutralization, reduced long-range repulsions, and efficient counterion-mediated bridging of distant monomers. This mechanism underlies the observed super-ballistic collapse dynamics ($\nu_c < 1$), emphasizing the inadequacy of neutral polymer models for describing the dynamics of inherently charged biopolymers like proteins and nucleic acids.

\begin{acknowledgments}
The simulations were carried out on the high performance computing machines Nandadevi at the Institute of Mathematical Sciences. SV is grateful to R Rajesh for critical discussions and inputs.
\end{acknowledgments}

%

\end{document}